\documentclass[aps,showpacs,floats,bibnotes,twocolumn]{revtex4}
\usepackage{epsfig,amsmath,xspace,dcolumn}

\newcommand{\EL}[2][]{{^{#1}{\rm{#2}}}}
\newcommand{\hag}{${^3{\rm H}(\alpha,\gamma)^7{\rm Li}}$}
\newcommand{\heag}{${^3{\rm He}(\alpha,\gamma)^7{\rm Be}}$}

\begin{document}

\title{Primordial nucleosynthesis with a varying fine structure
constant: An improved estimate}

\date{\today}

\author{Kenneth M. Nollett} 
\affiliation{Mail Code 130-33, California
Institute of Technology, Pasadena, CA\ \ 91125}

\author{Robert E. Lopez}
\affiliation{DAMTP, University of Cambridge, Cambridge CB3 0WA, England}

\begin{abstract}

We compute primordial light-element abundances for cases with fine
structure constant $\alpha$ different from the present value,
including many sources of $\alpha$ dependence neglected in previous
calculations.  Specifically, we consider contributions arising from
Coulomb barrier penetration, photon coupling to nuclear currents, and
the electromagnetic components of nuclear masses.  We find the
primordial abundances to depend more weakly on $\alpha$ than
previously estimated, by up to a factor of 2 in the case of $^7$Li.
We discuss the constraints on variations in $\alpha$ from the
individual abundance measurements and the uncertainties affecting
these constraints.  While the present best measurements of primordial
D/H, $^4$He/H, and $^7$Li/H may be reconciled pairwise by adjusting
$\alpha$ and the universal baryon density, no value of $\alpha$ allows
all three to be accommodated simultaneously without consideration of
systematic error.  The combination of measured abundances with
observations of acoustic peaks in the cosmic microwave background
favors no change in $\alpha$ within the uncertainties.

\end{abstract}


\maketitle

\section{Introduction}

The recent literature contains claims for time variation in the fine
structure constant $\alpha$ at a level of one part in $10^5$ since a
redshift of 2, based on observations of metal ions backlit by distant
quasars \cite{webb99,murphy,murphy01b,webb01}.  Regardless of
theoretical prejuduce for or against these claims
\cite{donoghue,banks}, it is of interest to examine other constraints
on $\alpha$ variation.  In particular, changing the fine structure
constant would influence the abundances of light nuclides produced
during big-bang nucleosynthesis (BBN).  BBN took place at redshifts
$10^8 \le z \le 10^{10}$, ending minutes after the big bang, and it is
the earliest cosmological test for which the basic physical laws are
well-understood.  The physics is less certain in varying-$\alpha$
scenarios, but BBN remains the earliest reasonable probe of variations
in $\alpha$.  It thus provides a large lever arm in the sense that if
$\alpha$ varies with time, larger variations may plausibly (though not
necessarily \cite{bekenstein,barrowetal}) be observed over longer
times.

Past BBN calculations for varying-$\alpha$ models focused on the
abundance of $\EL[4]{He}$~\cite{Kolb86,Barrow87,campbellolive}.  In
some ways, this keeps the physics simple, because $\alpha$ should
affect the $^4$He mass fraction $Y_P$ only through the neutron/proton
ratio at freezeout of weak interactions before BBN -- any effect on
the rates of subsequent nuclear reactions is irrelevant because
essentially all neutrons are gathered into $^4$He nuclei by the end of
BBN.  The likely size of the effect on $Y_P$ is also large compared
with observational errors if $|\delta|>0.05$, where
$\delta\equiv(\alpha-\alpha_0)/\alpha_0$ and subscript zero denotes
the present value of a quantity.

More recently, Bergstr{\"o}m, Iguri and Rubinstein~\cite{Bergstrom99}
(hereafter BIR) extended the varying-alpha calculation to include
other light nuclides by estimating the dependence of the
charged-particle reaction rates on Coulomb barrier penetration and
hence on $\alpha$.  However, they chose to neglect several places
where $\alpha$ enters the rates (from both barrier penetration and
other considerations), under the assumption that the exponential
$\alpha$ dependence in the Gamow factor will dominate all other
$\alpha$ dependence.  Our work corrects these omissions, most
importantly those arising from barrier penetration and from coupling
of photons to nuclear currents.  We also estimate the effects of
final-state barrier penetration and the $\alpha$ dependence of the
nuclear masses.  We use more elaborate models for the $\alpha$
dependences of the mass-7 radiative captures, and we correct BIR's use
of reaction rates that were already out of date when they did their
work.  

Finally, we elaborate the BIR discussion of BBN constraints on
$\alpha$, examining the region of the $(\Omega_B h^2,\alpha)$ plane
allowed by each of the observed light-nuclide abundances.  ($\Omega_B$
is the universal baryon density in units of the critical density; $h$
is the Hubble constant in units of $100\ \mathrm{km\ s^{-1}/Mpc}$.)
We then discuss the prospects for producing a consistent solution,
given systematic uncertainties in the observed abundances.  We leave
aside the question of how (or whether) other parameters of the
standard model of particle physics vary as $\alpha$ varies
\cite{calmet02,langacker,dent,Kolb86,campbellolive,beane02}, though
this may also affect BBN.

\section{Light nuclei and the fine structure constant}
\label{sec:effects}

The BBN light-element abundances, apart from $^4$He, depend on the
fine structure constant through nuclear reaction rates.  Calculating
how these rates depend on $\alpha$ is the main task undertaken by BIR.

Charged particles must tunnel through a Coulomb barrier to react at
low energies, and this produces most of the energy dependence in the
cross section.  Changing $\alpha$ raises or lowers the Coulomb
barrier, altering the reaction rates.  To describe laboratory results
and compute thermally-averaged rates, it is useful to separate out the
Coulomb part of the energy dependence by writing the low-energy cross
section as
\begin{equation}
\label{eqn:sfactor}
\sigma(E) = \frac{S(E)}{E} \, e^{-2\pi\eta} \,.
\end{equation}
Here, $\eta = \alpha Z_1 Z_2 \sqrt{\mu/2 E}$, $\mu$ is the reduced
mass, and $Z_1$, $Z_2$ are the charges of the incoming nuclei.  The
astrophysical $S$-factor $S(E)$ is then usually a slowly-varying
function of energy.

BIR calculate the $\alpha$ dependences of the thermal reaction rates
that result if the $S$-factors are assumed to be independent of
$\alpha$; \textit{i.e.,} if all of the $\alpha$ dependence is expressed
in Eqn. \ref{eqn:sfactor} and the definition of $\eta$.  We have
verified that their results are correct, given this assumption.
However, Eqn. \ref{eqn:sfactor} does not explicitly contain all of the
expected $\alpha$ dependence, even if the only effect of changing
alpha is to change the Coulomb barrier.  Moreover, BBN takes place at
relatively high energies (a few hundred keV) and $Z_1Z_2$ is only in
the range 1--4 in BBN, so the exponential does not dominate
sufficiently to make $\alpha$ dependence of the $S$-factor completely
negligible by comparison.  If BBN involved energies a factor of two
lower or nuclei with only slightly larger charges, corrections to the
BIR treatment would be negligible.

The first source of $\alpha$ dependence in $S(E)$ is Coulomb barrier
penetration.  Cross sections are proportional to the barrier
penetration factor $v_l$:
\begin{equation}
\label{eqn:penetration}
v_l = \frac{1}{F^2_l(\eta,kR)+G^2_l(\eta,kR)}.
\end{equation}
Here, $F_l$ and $G_l$ are the regular and irregular Coulomb wave
functions, respectively \cite{abramowitz}, $l$ is orbital angular
momentum of the entrance channel, $k$ is the wavenumber in the
entrance channel, and $R$ is the nuclear interaction radius.  (A clear
derivation of Eqn. \ref{eqn:penetration} may be found in Chapter VIII
of Blatt and Weisskopf \cite{blatt52}).  Assuming $l=0$ (a reasonable
assumption at very low energies), $v$ is nonzero even when $R=0$.  For
small $k$, it is reasonable to approximate the barrier-penetration
part of the cross section as
\begin{eqnarray}
\sigma(E) &\propto& v_l
\nonumber\\
&=& [G_0(\eta,0)]^{-2}
\label{eqn:R=0}
\\
&=&\frac{2\pi\eta}{e^{2\pi\eta}-1} 
\nonumber\\
  & \approx & 2 \pi \alpha Z_1Z_2 \sqrt{\mu/2E} \, e^{-2\pi\eta}
\,.
\label{eqn:G0}
\end{eqnarray}
We have thus ``derived'' the heuristic expression in
Eqn. \ref{eqn:sfactor}, and found that $S(E)$ is proportional to
$\alpha$.  If the entrance channels of all reactions are reasonably
approximated as $S$ waves penetrating to zero separation, it follows
that the charged-particle rates in BIR should be multiplied by an
additional factor of $\left(1+\delta\right)=\alpha_{\mathrm{BBN}}
/\alpha_0$ (where $\alpha_{\mathrm{BBN}}$ is the value of $\alpha$ at
the time of BBN).  When $R>0$ or $l>0$, Eqn.  \ref{eqn:G0} almost
always provides an overestimate of the sensitivity of
Eqn. \ref{eqn:penetration} to $\alpha$.  (Exceptions with $\sim 10\%$
underestimates of $dv_l/d\alpha$ occur at $E>100$ keV for reactions
with $Z_1Z_2=1$.)

The pedagogical literature usually motivates Eqn. \ref{eqn:sfactor} by
a WKB approximation.  Even where WKB provides a quantitatively good
approximation (not the usual case~\cite{blatt52}), the factor of
$\alpha$ on the front of Eqn. \ref{eqn:G0} enters the WKB barrier
penetration factor through the normalization that matches solutions
across the classical turning point.

When a reaction produces two charged particles, they must escape from
a Coulomb barrier, and this introduces more $\alpha$ dependence in the
rates.  The cross section is again suppressed by a factor of the form
shown in Eqn. \ref{eqn:penetration}, but with $E\rightarrow Q+E$ and
appropriate charges.  The effect is weak because $Q$ (the energy
released) is generally much larger than the Coulomb barrier, but there
are two exceptions, $^3{\rm He}(n,p)^3{\rm H}$ and $^7{\rm
Be}(n,p)^7{\rm Li}$.  These two cases are very interesting, because
their $Q$ values depend strongly on $\alpha$ (see below), and this
introduces $\alpha$ dependence in the final-state barrier penetration
that almost cancels the dependence due to the changing barrier height.

The generally high $Q$-values in BBN invalidate the approximation in
Eqn. \ref{eqn:G0} for final states.  In fact, the final-state barrier
penetration factors for the BBN reactions are almost independent of
energy over the BBN energy range.  Even Eqn. \ref{eqn:R=0} badly
overestimates the effect because $k$ is larger for the final states.
We estimate the final-state suppression by assuming $R=2$ fm and
$l=0,1,2$, computing logarithmic derivatives of $v_l$ with respect to
alpha, and looking for the largest effect.  The results should be
rough but reasonable upper limits to the size of the effect.  To keep
thermal rates simple, we exploit the weak energy dependence of these
derivatives and multiply the rates by energy-independent factors
$\left(1+ a\,\delta\right)$ which model their $\alpha$ dependence.  In
these factors, $a$ is estimated to be slightly larger in magnitude
than $\alpha v_l^{-1}\, dv_l/d\alpha$ for all $l$ considered, and the
adopted values are shown in Table \ref{tb:a}.  In the cases of $^3{\rm
He}(n,p)^3{\rm H}$ and $^7{\rm Be}(n,p)^7{\rm Li}$, the low $Q$-value
leads us to include in $a$ the complete dependence of final-state
channel widths on $Q$.  (Channel widths are proportional to the
product of channel velocity and $v_l$ \cite{blatt52}, both of which
vary with $Q$.)  This adjustment is based on our estimate of the
electromagnetic contribution to the masses (see below) and takes the
corresponding numbers in Table~\ref{tb:a} from slightly negative to
slightly positive.

\begin{table}
\begin{center}
\begin{ruledtabular}
\begin{tabular}{cd} 
Reaction & a \\
\hline 
$d(d,p)$ & -0.16 \\
$\EL[3]{He}(n,p)$ & +0.30 \\
$\EL[3]{He}(d,p)$ & -0.09 \\
$\EL[7]{Li}(p,\alpha)$ & -0.18 \\
$\EL[7]{Be}(n,p)$ & +0.20 \\
\end{tabular}
\end{ruledtabular}
\end{center}
\caption{Corrections for final-state interactions for processes
involving two charged particles in the final state. To correct the
rates, multiply by $\left(1+a\,\delta\right)$, where a is given
above.}
\label{tb:a}
\end{table}

The matrix elements for radiative captures -- photon-emitting
processes -- are proportional to $\alpha$ because $\alpha$ is the
strength of the coupling between the photon and nuclear currents.
Therefore, all radiative capture rates should be multiplied by an
additional factor of $\left(1+\delta\right)$.

Some of the radiative captures [$^3{\rm H}(\alpha,\gamma)^7{\rm Li}$
and $^3{\rm He}(\alpha,\gamma)^7{\rm Be}$] are also \textit{external}
reactions (e.g. \cite{christyduck,nollett2001}), so particles need not
penetrate the Coulomb barrier all the way to the nuclear surface; the
reaction matrix elements contain large contributions from outside the
nuclear interaction radius.  Further, Coulomb forces govern the forms
of the final states ($^7{\rm Li}$ and $^7{\rm Be}$ bound states) at
these distances.  Most importantly, the electrostatic energies of the
bound states are also large fractions of the reaction $Q$ values, so
that variations in $\alpha$ change the outgoing photon energy
significantly.  This energy enters the cross sections as
$\sigma\propto E_\gamma^3$.  We used a simple radiative capture model
to estimate the combined size of these three effects, though we found
that the last one dominates.

We assume the following: 1) These reactions can be treated as direct
captures in which the bound states consist of smaller clusters:
$\EL[4]{He}$ and $\EL[3]{H}$ in the case of $\EL[7]{Li}$, or
$\EL[4]{He}$ and $\EL[3]{He}$ in the case of $\EL[7]{Be}$.  2) Wave
functions can be generated from potential models; we use the potential
of Buck et al. \cite{buck85}.  3) The only effect of changing $\alpha$
is to change the charges in the Coulomb term of this potential.  There
is no question that the assumptions numbered 1 and 2 are reasonable
descriptions of laboratory data.  It is plausible that assumption
number 3 is no worse than other assumptions we have made implicitly in
modelling BBN with varying $\alpha$.  We estimate the variation of the
thermal reaction rates with $\alpha$ from this model.  We find the
effect on the thermal rates (beyond the effect in
Eqn. \ref{eqn:sfactor}) to be roughly independent of temperature, and
we characterize it by fits to the computed variation \textit{beyond} the
corrections already discussed,
\begin{equation}
f=\left(\frac{\alpha}{\alpha_0}\right)^{-2}\frac{\langle\sigma(\alpha)v\rangle}
{\langle\sigma(\alpha_0)v\rangle}
\label{eqn:capture}
\end{equation}
(with $\langle\sigma v\rangle$ the thermally-averaged rate).  These
fits give estimates of the rates at $\alpha\neq\alpha_0$ that are
accurate to 20\% in the \textit{additional} effect beyond the exponential
factor and two powers of $\alpha$ already considered (at temperatures
less than $10^{10}$ K), and they are shown in Table \ref{tab:capture}.
We assume that this $\alpha$ dependence holds relative to the reaction
rates computed from laboratory measurements.  (See
Sec. \ref{sec:rates} for the baseline $\alpha=0$ rates.)

\begin{table}
\begin{ruledtabular}
\begin{tabular}{lcccccc} 
Reaction & $a_0$ & $a_1$ & $a_2$ & $a_3$ & $a_4$ & $a_5$ \\
\hline 
$^3{\rm H}(\alpha,\gamma)^7{\rm Li}$  &  1 & 1.3718 &   0.502 &  0.1829 & 0.2693 &  $-0.2182$ \\
$^3{\rm He}(\alpha,\gamma)^7{\rm Be}$ &  1 & 2.148  &   0.6692 &  $-5.566$ &  
$-10.63$ & $- 5.730 $ \\
\end{tabular}
\end{ruledtabular}
\caption{Estimated dependence of radiative capture rates on $\alpha$,
with $f(\alpha)=\sum a_i \delta^i$ as defined in
Eqn. \protect\ref{eqn:capture}.}
\label{tab:capture}
\end{table}

Nuclear masses depend on $\alpha$ through electromagnetic interactions
between nucleons.  In turn, differences between initial-state and
final-state masses are used in nucleosynthesis calculations to
determine the rates of reverse reactions (by detailed balance).  BIR
report that reverse reactions have essentially no effect on the BBN
yields, but we find that when we alter the masses, the yields change
at $\Omega_Bh^2>0.025$.

We adopt electromagnetic contributions $M_{EM}$ to nuclear masses
based on quantum Monte Carlo calculations \cite{pudliner,wiringa},
where all electromagnetic contributions (including magnetic-moment
interactions) are computed explicitly.  We then multiply $M_{EM}$ by
$\left(1+\delta\right)$ to vary $\alpha$, and adjust the Q-values
accordingly to obtain $Q = Q_0 + \Delta M_{EM}\delta$. The adopted
values of $\Delta M_{EM}$ are given in Table~\ref{tb:q}.

\begin{table}
\caption{Electromagnetic contributions $\Delta M_{EM}$ to the reaction
energy yields, from the calculations of
Refs. \protect\cite{pudliner,wiringa}.  Shown for comparison are the
total energy yields $Q_0$.}
\label{tb:q}
\begin{ruledtabular}
\begin{tabular}{cdd} 
{Reaction} & {\Delta M_{EM}\ \mathrm{(MeV)}} & {Q_0\ \mathrm{(MeV)}} \\
\hline
$p(n,\gamma)d$ &  -0.018 & 2.224\\
$d(p,\gamma)\EL[3]{He}$ & -0.70 & 5.4940\\
$d(d,p)t$ & -0.004 & 4.033\\
$d(d,n)\EL[3]{He}$ & -0.682 & 3.270 \\
$t(d,n)\EL[4]{He}$ & -0.81 & 17.590\\
$\EL[3]{He}(d,p)\EL[4]{He}$ & -0.81 & 18.353\\
$t(\alpha,\gamma)\EL[7]{Li}$ & -0.844 & 2.467\\
$\EL[3]{He}(\alpha,\gamma)\EL[7]{Be}$ & -1.74 & 1.587\\
$\EL[3]{He}(n,p)\EL[3]{H}$ &\ \ 0.678 & 0.7637\\
$\EL[7]{Li}(p,\alpha)\EL[4]{He}$ & \ \ 0.016 & 17.3459\\
$\EL[7]{Be}(n,p)\EL[7]{Li}$ & \ \ 1.574 & 1.6443 \\
\end{tabular}
\end{ruledtabular}
\end{table}

In summary, to calculate BBN yields for an altered fine structure
constant, first apply the changes described in BIR. Then apply these
additional corrections:
\begin{itemize}
\item Coulomb normalization factor: Multiply all charged-particle
rates by a factor of $\left(1+\delta\right)$.
\item Radiative captures: Multiply rates for radiative captures by
another factor of $\left(1+\delta\right)$.  Multiply the rates for
\heag\ and \hag\ by the additional factors $f$ in Table
\ref{tab:capture}.
\item Final-state Coulomb interactions: Multiply rates with two
charged particles in the final state by a factor
$\left(1+a\,\delta\right)$, using the values of $a$ given for each
reaction in Table ~\ref{tb:a}.
\item Masses: Adjust the reaction $Q$-values using the electromagnetic
contributions shown in Table~\ref{tb:q}: $Q=Q_0+\Delta M_{EM}\delta$.
\end{itemize}

Most of these effects have the opposite sign from the exponential
factor considered in BIR.  We consequently expect a weaker dependence
on $\alpha$ for the reaction rates and resulting BBN yields.

The uncertainty in this improved treatment is difficult to assess.
The adjustment of the $Q$-values is relatively model-independent.
Adjustment of the rates, on the other hand, is less certain because we
have used approximations to the Coulomb penetrabilities, and also
because we have neglected everything happening inside the nucleus.
The result is certainly better (and closer to self-consistency) than
the overestimate that results from assuming that $S$-factors are
independent of $\alpha$.  Our approximations to barrier
penetrabilities almost always overestimate the effect of varying
$\alpha$.  The overestimates are generally 5--10\% of the variation,
except where $Z_1Z_2=1$; here, they may be underestimates by up to
10\% of the variation.  These percentages are relative to barrier
penetration factors for $R=2$ fm and $l<2$, and not relative to the
(unknown and perhaps unknowable) total change in the cross section
when $\alpha$ changes.  The most extreme limit to the validity of the
method is where states become unbound, i.e, $Q=Q_0+\Delta M_{EM}\delta
= 0$.  This occurs at about $\delta=0.5$ for the first excited state
of $^7$Be, although our radiative capture calculation fails for
numerical reasons beyond about $\delta=0.3$.  Our treatment of
final-state barrier penetration also assumes that $\delta$ is
``small'' (being based on first derivatives), but these corrections
turn out to be relatively unimportant.  We assume symmetric limits to
the validity of our analysis, $-0.3 < \delta < 0.3$, with
less-questionable validity at $-0.1 < \delta < 0.1$.

\section{Laboratory Rates}
\label{sec:rates}

We note finally that BIR used reaction rates that were already out of
date, and they did not cite sources for the rates they did use.  All
of their important rates come from the evaluation by Smith, Kawano,
and Malaney (SKM) \cite{skm} (and the others trace to the Caughlan and
Fowler compilations \cite{cf88}).  In the time since SKM (1993), some
of the most uncertain cross sections have been re-measured.  Newer,
better rates exist for the processes
$d(p,\gamma)\EL[3]{He}$~\cite{dpg-schmid,dpg-ma},
$t(\alpha,\gamma)\EL[7]{Li}$ \cite{tag-brune}, and
$\EL[3]{He}(n,p)\EL[3]{H}$ \cite{he3np-brune}.  The revised rate for
the first reaction has the usual functional form derived by the
$S$-factor formalism (though its parameters are actually fitted to a
numerical integration), and the expression from BIR, as corrected in
Sec. \ref{sec:effects}, may plausibly provide its $\alpha$ dependence.
We have thus substituted the coefficients from Ref.  \cite{dpg-ma}
into the BIR expression.

For the second reaction, the revised rate is again a fit to a
numerical integration, but the functional form used does not show up
in the BIR calculation.  Since we have a physical model for the
variation of this rate with $\alpha$, and the BIR formalism does in
fact account for much of the $\alpha$ dependence in this model, we
adopt the following somewhat convoluted procedure: we first computed a
fitting formula for the ratio of the revised rate to the SKM rate.  In
our BBN code, we multiply the BIR $\alpha$-dependent rate by this
fitting formula.  We then apply the corrections from
Sec. \ref{sec:effects}.  The result provides both the correct,
up-to-date rate at $\alpha=\alpha_0$, and the $\alpha$ dependence
computed from our direct capture model.  The SKM $S$-factor contains a
term of the form
\begin{equation}
S(E)=S(0)q(E)e^{-\beta E},
\end{equation}
which we drop in this procedure to more accurately reproduce the
$\alpha$ dependence computed in our model.  This is not a problem with
the SKM rate evaluation, but rather a question of producing simple,
adequate fitting functions for the $\alpha$ dependence of the rate.
(Note that in principle, this term arises from the external-capture
nature of the reaction, so that $\beta$ should almost certainly depend
on $\alpha$ -- a possibility not considered in BIR.)

Finally, we updated the rate for $\EL[3]{He}(n,p)\EL[3]{H}$, which
does not need the BIR treatment because of the neutron in the entrance
channel.  In summary, our calculations used the SKM rates, with
$\alpha$ variation estimated as described above, and with updates from
Refs. \cite{dpg-ma,tag-brune,he3np-brune}.  We compared a calculation
using these rates and $\alpha_{\mathrm{BBN}}=\alpha_0$ with the
corresponding results from the NACRE~\cite{nacre} and Nollett \& Burles
\cite{nb2k} evaluations of the rates (which are not suited to
computing $\alpha$ variation).  The results were in agreement within
the uncertainties.

To compare our calculation with observed abundances, it will be
necessary to take uncertainties on the laboratory measurements into
account.  The effects of these uncertainties on the computed
abundances can, in general, only be estimated by Monte Carlo methods.
Since we compute a large grid of models in $\Omega_B h^2$ and
$\alpha$, such a procedure would be time-consuming.  Instead, we make
very simple, conservative estimates of the uncertainties.  The
uncertainties on the computed helium mass fraction $Y_P$ have been
estimated by Lopez \& Turner~\cite{lopezturner} to be $\pm 0.0004$
from experiment and $\pm 0.0002$ from theory.  We have not included
their elaborate treatment of small effects in our calculation, and we
will assume an error of $0.0005$ on the calculation.  For
$\Omega_Bh^2>0.011$, the uncertainty in D/H arises mainly from the
rate for $d(p,\gamma)^3{\rm He}$; we use an estimate of 9\% here,
based on the uncertainty in D/H at $\Omega_B h^2=0.036$, its maximum
in Ref. \cite{nb2k}.  Similarly, the uncertainty in $^7$Li/H at
$\Omega_Bh^2> 0.011$ is governed by the uncertainty in the rate for
\heag.  We take as the uncertainty in $^7$Li/H its maximum of 0.05 dex
at $\delta=0$ from Ref. \cite{nb2k}.  We will see below that $^3$He/H
depends only weakly on $\alpha_\mathrm{BBN}$.  Moreover, the
identification of the observed $^3$He/H with the primordial value
\cite{rood} seems to us problematic, so we will not use $^3$He in our
examination of the evidence for varying $\alpha$.

\section{Results}

\begin{figure}
\centerline{\epsfig{file=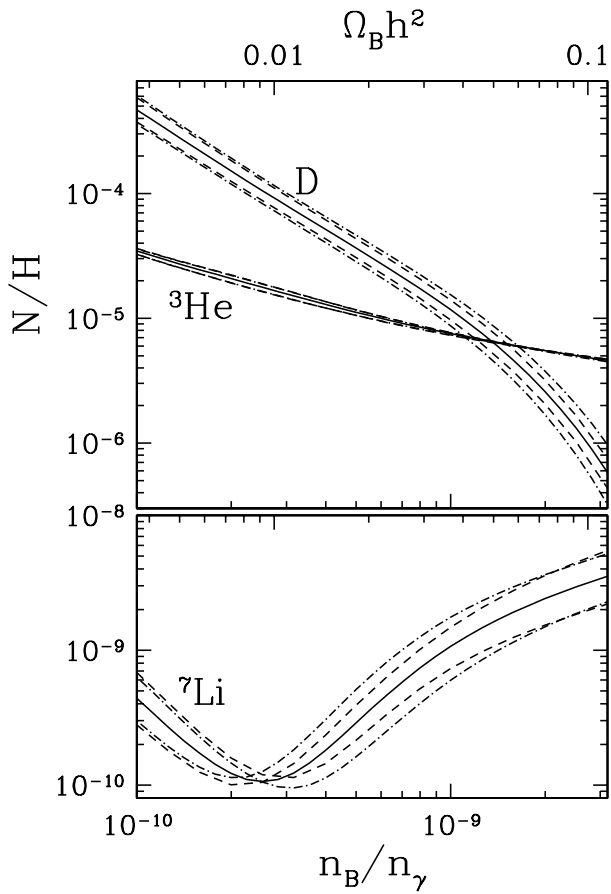,width=3.5in,angle=-0}}
\caption{Light element abundances for $\delta = 0$ (solid lines),
$\delta=\pm 0.05$ from the BIR method (dash-dotted lines) and
$\delta=\pm 0.05$ from our calculation (dashed lines).  $n_B$ and
$n_\gamma$ are the number densities of baryons and photons,
respectively.}
\label{fg:bbn}
\end{figure}

We incorporated all of the $\alpha$ dependence discussed above into
the standard BBN code~\cite{kawano,kawano2} and computed yields.
Figure~\ref{fg:bbn} shows the resulting light-element abundances for
$\delta = 0$, $\pm 0.05$, compared with a calculation applying the BIR
formalism to updated laboratory rates.  For D/H and $^7$Li/H, we find
that the $\alpha$ dependence is weaker than that found by BIR, though
the effect still works in the same direction.  This can be understood
as follows: the strongest $\alpha$ dependence in their calculation is
in the suppression of charged-particle rates by the negative
exponential in those rates.  In our calculation, positive powers of
$\alpha$ from current coupling and careful treatment of the barrier
penetration partially cancel this suppression.  The $^7$Li also
receives a somewhat smaller contribution from adjustments in outgoing
photon energy which we modelled.  Effects of the changing $Q$ values
on the reverse reactions become important for $^7$Li/H at large
$\Omega_B h^2$, making our estimate coincidentally equal to that of
BIR at $\Omega_B h^2 > 0.065$.  Our estimate of final-state effects
makes a much smaller contribution to the yields than any of the other
corrections we have added to the calculation.

\begin{figure}
\centerline{\epsfig{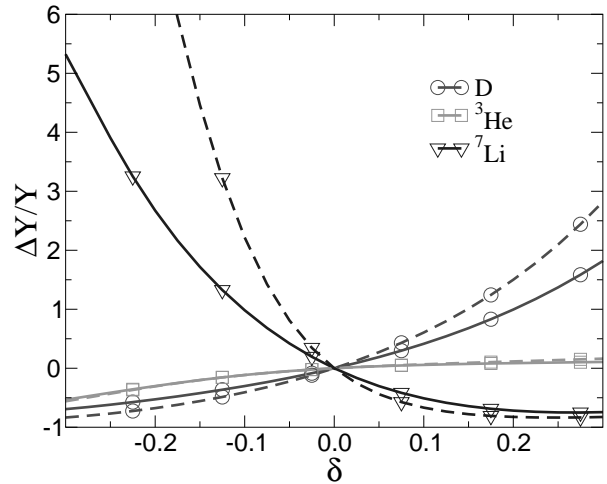}}
\caption{Fractional change in yields from their value at $\delta=0$,
computed at $n_B/n_\gamma=5.6\times10^{-10}$ ($\Omega_Bh^2=0.0204$).
Solid curves are from our calculation.  Dashed curves indicate results
from the methods of BIR.}
\label{fig:breakdown}
\end{figure}

Our results for the $\EL[4]{He}$ abundance agree with BIR. This
abundance depends on $\alpha$ primarily through the neutron-proton
mass difference $m_n-m_p$ which governs the neutron/proton ratio at
the start of BBN.  Like BIR, we use the phenomenological prescription
of Gasser and Leutwyler~\cite{Gasser82} to model the $\alpha$
dependence of the nucleon mass splitting:
\begin{equation}
m_n-m_p = 1.3 \left( 1+ c\,\delta \right) \, \rm{MeV} \,,
\label{eq:q}
\end{equation}
where $c = -0.59$.  As pointed out in BIR, this model is plausible,
but certainly not definitive, so the $\EL[4]{He}$ yields are less
certain than those of other light nuclides.  We use the Gasser \&
Leutwyler value because we did not find competing, explicit values of
$c$ in the literature.  The abundances of the other nuclides are for
the most part affected only weakly by this uncertainty, but the
nucleon mass splitting makes an important contribution to the $\alpha$
dependence of $^7$Li/H at low $\Omega_B h^2 \le 0.010$.  We show the
dependences of D/H, $^3$He/H, and $^7$Li/H on $\delta$ at
$\Omega_Bh^2=0.0204$ in Fig. \ref{fig:breakdown}.

\section{Observational constraints}

Given observationally-inferred primordial abundances, our calculations
can be used to constrain possible combinations of $\Omega_Bh^2$ and
$\delta$.  In principle, two abundances suffice to fix values of the
two parameters, and a third abundance provides a check on the result.
In reality, there are systematic errors associated with both the
measurements and the varying-$\alpha$ calculation, and we find no
solution which satisfies the D, $^4$He, and $^7$Li observations
simultaneously.  ($^3$He/H shows relatively weak dependence on
$\alpha$; it may in principle help constrain $\Omega_Bh^2$ so that
$\alpha$ may be extracted from other abundances, given the primordial
ratio.)  We now examine the constraints provided by the present
observations.  This discussion is in large part supplementary to those
in BIR and in Avelino et al. \cite{Avelino01} (who used the BIR rates,
but also applied statistical error estimation to a subset of the
data).  However, we also draw upon our improved estimation of the
yields.  The reader should keep in mind that the primordial
light-nuclide abundances require difficult observational work and are
by no means settled questions; we will attempt to point out the most
serious problems, but the reader is referred to recent reviews of the
topic for critical discussion of the uncertainties and appropriate
\textit{caveats} \cite{dns_vol,tytler-review}.  It what follows, we
assume Gaussian distributions for all uncertainties and include the
roughly-estimated yield uncertainties of Sec. \ref{sec:rates}.

\begin{figure}
\centerline{\epsfig{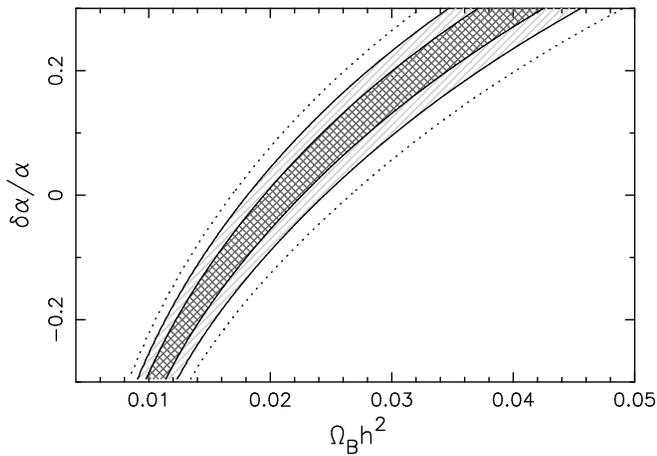}}
\caption{Regions of $1\sigma,\ 2\sigma,$ and $3\sigma$ confidence,
based on the D/H measurement of O'Meara et al.\protect\cite{omeara}.}
\label{fig:d}
\end{figure}

A value for a single light-nuclide abundance specifies one constraint
relating $\Omega_Bh^2$ and $\delta$, and therefore defines a curve in
the $(\Omega_Bh^2,\delta)$ plane.  As a first example, we show in Fig.
\ref{fig:d} the constraint in the $(\Omega_Bh^2,\delta)$ plane
provided by an average of the extragalactic D/H measurements of Tytler
and collaborators, $(3.0\pm 0.4) \times 10^{-5}$~\cite{omeara}.
(There exist measurements by other groups, which we exclude because of
concerns about signal-to-noise \cite{pettini} and complex velocity
structure \cite{dodorico,levshakov}; these tend to increase the
scatter.)  The largest weakness of D/H as a probe of BBN is the
difficulty of obtaining large enough numbers of high-quality data to
be confident in their analysis.  Because there is uncertainty in the
observations, there is uncertainty in specifying the curve; this is
indicated by shaded regions delimiting 1, 2, and $3\sigma$ confidence
levels.  We note that the error estimate of Ref. \cite{omeara} is
rather conservative.

\begin{figure}
\centerline{\epsfig{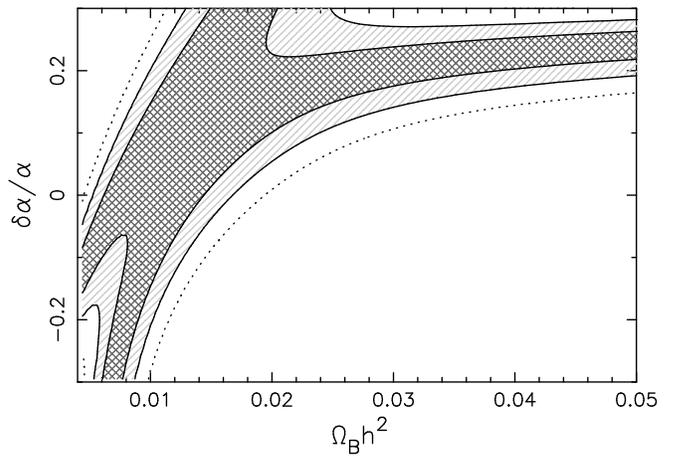}}
\caption{Regions of $1\sigma,\ 2\sigma,$ and $3\sigma$ confidence,
based on the $^7$Li/H value from Ryan \protect\cite{ryan00}.}
\label{fig:7li}
\end{figure}

\begin{figure}
\centerline{\epsfig{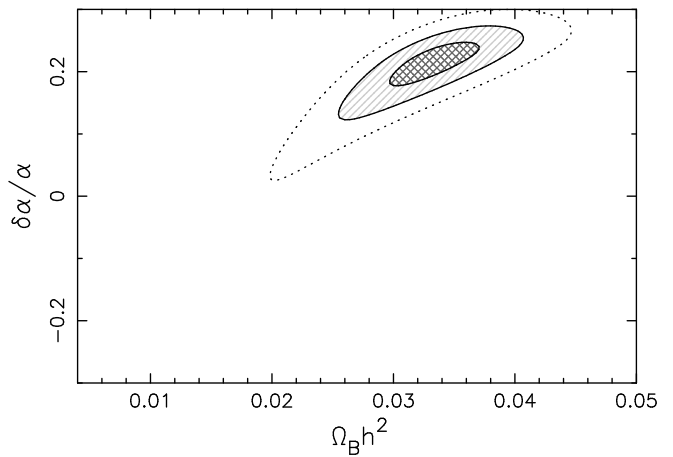}}
\caption{Confidence levels in the $\left(\Omega_Bh^2,\delta\right)$
plane for D/H and $^7$Li/H combined, ignoring $^4$He. The contours
reflect $1\sigma$, $2\sigma$, and $3\sigma$ confidence levels. The
solution for these two parameters given the two constraints is
$\Omega_B h^2=0.034$, $\delta = 0.23$, but the result fails a check
against a third constraint.}
\label{fig:d+7li}
\end{figure}

The analogous curves based on halo-star observations of $^7$Li are
shown in Fig. \ref{fig:7li}.  We derive these curves from the
primordial $^7$Li abundance advocated by Ryan \cite{ryan00} and Ryan
et al.~\cite{ryanetal00}, $\log_{10}(^7{\rm
Li/H})=-9.91^{+0.19}_{-0.13}$.  It should be kept in mind that there
could be factor-of-two effects associated with Li destruction in the
observed stars \cite{vauclair,pinsonneault,salaris} and with the
stellar atmosphere models used to derive abundances \cite{asplund}.
There are two curves for $^7$Li/H, reflecting the two paths for its
production in BBN: a low-$\Omega_B$ case in which $^7$Li is produced
directly, and a high-$\Omega_B$ case in which $^7$Li is produced as
$^7$Be, followed by beta decay.  Again, confidence intervals are
indicated by the shaded regions.  Comparison of Figs. \ref{fig:d} and
\ref{fig:7li} shows one point where the D/H and $^7$Li results can be
reconciled.  This point occurs at $\Omega_Bh^2=0.034$ and
$\delta=0.23$.  Although we consider the range of validity of our
treatment to be $-0.3 < \delta < 0.3$, there does appear to be a
second solution at lower $\Omega_Bh^2$ and an extreme negative
$\delta$.  Note that the best-fit model for D and $^7$Li with
$\delta=0$ is rejected at 98\% confidence.  The joint constraint is
shown in Fig. \ref{fig:d+7li}.

\begin{figure}
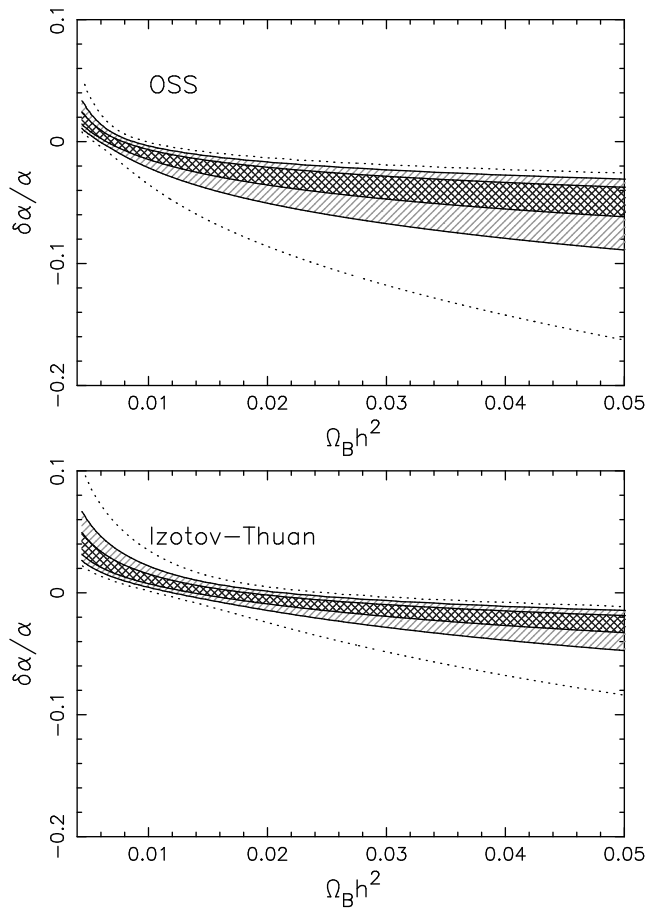

\centerline{\epsfig{file=conf4He-alt-restrict.w.ps,width=2.35in,angle=-90}}
\centerline{\epsfig{file=conf4He-restrict.w.ps,width=2.35in,angle=-90}}
\caption{Regions of $1\sigma,\ 2\sigma,$ and $3\sigma$ confidence,
based on the $Y_P$ value from OSS \protect\cite{OSS98} (top) and
IT \protect\cite{Izotov98} (bottom).}
\label{fig:4he}
\end{figure}

How seriously should this result be taken?  A solution can always be
found for two abundances [assuming that the curves in the
$(\Omega_Bh^2,\alpha)$ plane intersect], so its validity must be
checked with a third abundance.  The other abundance at our disposal
is that of $^4$He, as observed in extragalactic H{\sc ii} regions.
Here, we confront unavoidable systematic problems associated with the
extraction of abundances from spectral-line data.  The two largest
compilations of data yield discrepant values for the primordial $^4$He
mass fraction: Olive, Skillman, and Steigman (OSS) find $Y_P=0.234\pm
0.002$ \cite{OSS98} while Izotov and Thuan (IT) find $Y_P=0.244\pm
0.002$ \cite{Izotov98}.  The two resulting curves are shown in
Fig. \ref{fig:4he}.  It seems clear from these curves that \textit{if}
agreement could be reached concerning the value of $Y_P$, quite
stringent constraints could be placed on $\delta$ irrespective of
$\Omega_B$.  However, this impression is illusory because such
constraints involve the size of the electromagnetic contribution to
the nucleon mass splitting.  As discussed above, this number is
difficult to calculate, and subject to large uncertainty.  The
confidence intervals shown include allowances for a 30\% Gaussian
uncertainty in this parameter (the source of the asymmetric
intervals).

Comparing Figs. \ref{fig:d+7li} and \ref{fig:4he}, we see that the
D+$^7$Li solution above fails the consistency check with $^4$He.
Making this solution work would require either $Y_P$=0.354 --
substantially larger than even the solar value of 0.275 -- or else
$c=0.045$ (at $Y_P=0.244$) or $c=0.10$ (at $Y_P=0.234$), compared to
the current best guess of $c=-0.59$.  While $c$ is very uncertain, it
cannot be positive if the electromagnetic force contributes to the
baryon masses only through self-energies and interactions among
constituent quarks \cite{Gasser82}.  Put another way, a $\chi^2$ fit
including all three abundances gives a minimum $\chi^2$ far from the
D/H + $^7$Li solution (because of the small error bars on $Y_P$), but
the likelihood at this minimum is low: a fit using the IT $Y_P$ is
rejected at 97.5\% confidence (actually worse than if $\delta=0$ is
assumed, since the extra parameter removes a degree of freedom), while
a fit using the OSS $Y_P$ is rejected at 98.0\% confidence.  The exact
numerical values of these confidence limits are probably not
meaningful, since the $Y_P$ error bars in particular are smaller than
the systematic uncertainties, but the indication is in any case that
varying $\alpha$ does not describe the data significantly better than
leaving it fixed.

There is thus no solution favored by all three observed primordial
abundances, even at $\delta=0$.  Perhaps one of the pairwise
combinations of abundances provides an acceptable solution that
requires only plausible systematic problems with the third observed
abundance.  Because the D+$^7$Li solution requires such a stretch to
match the $^4$He observations, we can probably rule it out.  The
D+$^4$He case is interesting because it does not require an
``extreme'' value for $\delta$.  This is in the sense that existing
claims for time variation in $\alpha$ based on metal lines in quasar
absorption systems are on the order of $10^{-5}$.  The best fits for
BBN with D and $^4$He only are at $\Omega_Bh^2=0.020\pm 0.005$,
$\delta=-0.007^{+0.010}_{-0.017}$ (IT) and
$\Omega_Bh^2=0.020^{+0.005}_{-0.004}$, $\delta=-0.03^{+0.01}_{-0.04}$
(OSS; all errors $2\sigma$).  These solutions have the advantage of
agreeing with the value of $\Omega_Bh^2$ derived from observations of
the cosmic microwave background (CMB).  However, $^7$Li/H is
discrepant by a factor of 2--3, so this is no better than the standard
$\alpha_\mathrm{BBN}=\alpha_0$ case in its need for a greatly
underestimated primordial $^7$Li/H.  

The final pair of abundances available is $^4$He+$^7$Li.  Any $^4$He
constraint crosses both curves defined by the $^7$Li constraint, and
either of the two $^4$He data sets could be correct.  There are thus
four different solutions we could consider, but it should suffice to
say that they all favor $|\delta | < 0.02$ and $\Omega_B h^2 < 0.017$.
The solution least discrepant from the D/H data requires
D/H=$7.3\times 10^{-5}$, more than twice the current best
observational average and greater than the highest value presently
claimed.

Thus, given the present observational data for light-element
abundances, allowing $\alpha$ to vary does not allow BBN to fit the
data in a way that is convincingly better than in the simplest
$\alpha_\mathrm{BBN}=\alpha_0$ case.  The existing discrepancies
between the light-nuclide abundances and the detailed predictions of
BBN are better-explained in terms of systematic effects than in terms
of varying $\alpha$.  The above analysis produces two plausible
constraints on $\alpha_\mathrm{BBN}$: one based on deuterium that
derives its strength from the simplicity of the D/H measurement
(Fig. \ref{fig:d}), and one based on $^4$He that derives its strength
from the presumptive strong dependence of $Y_P$ on
$\alpha_\mathrm{BBN}$ (Fig. \ref{fig:4he}).  The constraint based on
$^4$He which was derived in BIR remains the best that one can do to
constrain $\alpha_\mathrm{BBN}$ independent of any other parameter,
with the \textit{caveat} that the nucleon mass splitting is not
well-understood, and so the constraint could be much weaker than it
appears.  The constraint based on D/H is perhaps more secure because
the varying-$\alpha$ D/H yields depend on relatively well-understood
physics and because the D/H measurement is in principle relatively
free of systematic pitfalls: it involves demonstrably pristine
material, and it does not depend on radiative transfer or ionization
corrections.  The deuterium constraint may be expressed as
$\log_{10}\Omega_B h^2-\delta = -1.68\pm0.07$ at $2\sigma$, but it has
the disadvantage that one can obtain large variations in $\alpha$
merely by adjusting $\Omega_B$.  Given the large regions of high
confidence level in Fig. \ref{fig:7li}, $^7$Li does not provide a
useful constraint by itself.

We note that the joint constraint from D/H and the IT value of $Y_P$
was also examined in Ref. \cite{Avelino01}.  These authors obtained a
result of $\delta=-0.007\pm 0.009$ (compared with our
$\delta=-0.007^{+0.010}_{-0.017}$ for this case), and identical limits
on $\Omega_Bh^2$ to those found here.  One would expect differing
values of $\delta$ between their analysis and ours: in our
calculation, D/H depends more weakly on $\delta$, and the updated
laboratory cross section for the crucial $d(p,\gamma)^3\mathrm{He}$
cross section is lower than what they used.  It appears that these two
effects have cancelled by chance at the best-fit solution.  ($^4$He
should behave the same way in both calculations, because we have used
the same value of $c$).  The differing error limits result from a
combination of the weaker dependence of $d(p,\gamma)^3\mathrm{He}$ on
$\delta$ in our treatment, the inclusion of a 9\% cross section
uncertainty on our predicted D/H, and the 30\% uncertainty that we
have assigned to $c$.

Finally, an independent constraint on $\Omega_Bh^2$ now exists from
measurements of CMB anisotropies.  The current results are
$\Omega_Bh^2=0.022\pm 0.003$ from BOOMERANG \cite{newboom},
$\Omega_Bh^2=0.022\,^{+0.004}_{-0.003}$ from DASI \cite{dasi}, and
$\Omega_Bh^2=0.033\pm 0.007$ from MAXIMA \cite{maxima} (all $1\sigma$
errors).  A single direct measurement of $\Omega_Bh^2$ is a vertical
line in Figs. \ref{fig:d}--\ref{fig:4he}.  Values of
$\alpha_\mathrm{BBN}$ based on these measurements and the individual
light-nuclide abundances are shown in Table \ref{tab:deltas}.  A firm
value of $\delta$ may not be derived from this table, because the
relative reliabilities of the inferred primordial abundances are not
known.  However, a general trend is clear: only the combination
$^7$Li/H + MAXIMA or the combination of OSS $^4$He with any of the CMB
data provides any strong indication of $\delta\neq 0$.  For all other
data, the $2\sigma$ limits either include $\delta=0$ or nearly include
it (missing by 10\% or less of the best-fit value).

In principle, the estimation of $\Omega_Bh^2$ from CMB measurements
also depends on the value of $\alpha$ at matter-radiation decoupling
\cite{Kaplinghat99,Hannestad99,Battye00}, $\alpha_\mathrm{CMB}$.  This
may be included among the parameters to be determined jointly from the
CMB, or some value may be assumed as a prior.  The results quoted
above and used to derive Table \ref{tab:deltas} are based on the
assumption that $\alpha_\mathrm{CMB}=\alpha_0$.  The values of
$\alpha_\mathrm{CMB}$ and $\alpha_\mathrm{BBN}$ may have nothing to do
with each other: $\alpha$ may evolve between these events.  However,
barring entropy increases after BBN, the same $\Omega_Bh^2$ today
should be inferred from both BBN and the CMB.  After marginalizing
over $\alpha$, Avelino et al. \cite{Avelino01} find from a combination
of the BOOMERANG, DASI, and COBE data that
$\Omega_Bh^2=0.019^{+0.004}_{-0.002}$ at $1\sigma$.  For each BBN
nuclide, Table \ref{tab:deltas} shows the effects of combining the BBN
constraint with this value of $\Omega_Bh^2$.  The results are not
markedly different than for the DASI and BOOMERANG results
independently.  This is not surprising, because Avelino et
al. \cite{Avelino01} find $\alpha_\mathrm{CMB}=\alpha_0$, within the
uncertainties.

\begin{table}
\caption{Constraints obtained by combinining individual light-element
abundances with $\Omega_Bh^2$ as inferred from three sets of data on
acoustic oscillations in the CMB, and from the analysis of Avelino et
al. \protect\cite{Avelino01}, which marginalized over $\alpha$.
Numbers shown are the best-fit $\delta$ and the $\delta$ at the $\pm
2\sigma$ boundaries.}
\label{tab:deltas}
\begin{ruledtabular}
\begin{tabular}{llddd} 
{Abundance} & {CMB experiment} & -2\sigma  & {\delta}   & +2\sigma\\
\hline
$^4$He (Izotov) & MAXIMA      & -0.053 & -0.017  & -0.0028 \\
                & DASI        & -0.031 & -0.0083 &  0.0045 \\
                & BOOMERANG   & -0.030 & -0.0083 &  0.0027 \\
                & Marginalized& -0.011 & -0.0052 &  0.0054 \\
\\			    	               
$^4$He (OSS)    & MAXIMA      & -0.12  & -0.039  & -0.020 \\
                & DASI        & -0.087 & -0.030  & -0.015 \\
                & BOOMERANG   & -0.086 & -0.030  & -0.016 \\
                & Marginalized& -0.078 & -0.027  & -0.014 \\
\\
D/H             & MAXIMA      & -0.051 &  0.22   &  0.3+\footnotemark[1] \\
                & DASI        & -0.18  &  0.027  &  0.20 \\
                & BOOMERANG   & -0.14  &  0.027  &  0.18 \\
                & Marginalized& -0.17  & -0.038  &  0.15 \\
\\			    	               
$^7$Li/H        & MAXIMA      &  0.094 &  0.22   &  0.3+\footnotemark[1] \\
                & DASI        & -0.003 &  0.18   &  0.3+\footnotemark[1] \\
                & BOOMERANG   &  0.022 &  0.18   &  0.3+\footnotemark[1] \\
                & Marginalized& -0.011 &  0.149  &  0.3+\footnotemark[1] \\
\footnotetext[1]{$2\sigma$ upper boundary exceeds validity of model.}
\end{tabular}
\end{ruledtabular}
\end{table}

\section{Conclusions}

We have improved the calculation of primordial abundances in
varying-$\alpha$ scenarios by including corrections previously
neglected: normalization of initial-state Coulomb penetrabilities,
final-state Coulomb interactions, photon coupling to nuclear currents,
photon energy and barrier penetration in external direct captures, and
electromagnetic contributions to the nuclear masses.  These affect the
interactions between composite nuclei, so they affect all of the BBN
abundances except that of $^4$He.  In sum, these corrections can
amount to half of the effect previously computed for $^7$Li/H and
about 1/3 of the previous effect on D/H.  The most important
corrections are the initial-state penetrability and current-coupling
contributions.  The nuclear mass modifications become important only
for $\Omega_Bh^2 > 0.036$.  BBN is less sensitive to $\alpha$ than
previous calculations indicated, but the sign of the dependence
remains the same.  

Applying our calculation to the data, the most secure constraint we
arrive at is $\log_{10}\Omega_B h^2 -\delta= - 1.68\pm 0.07$
($2\sigma$), based on the D/H measurements of Tytler and
collaborators~\cite{omeara}.  Combining this constraint with a
constraint $\Omega_Bh^2=0.022\pm 0.003$ from CMB anisotropies yields
$\delta=0.03\pm 0.07$ ($1\sigma$).  A separate constraint, subject to
both theoretical and astronomical uncertainties which are difficult to
assess, is provided by $^4$He measurements: the BIR result of
approximately $|\delta|< 0.1$ stands as the best that can be done
here, about the same as the D/H + CMB constraint.

Other recent work on time-varying $\alpha$ suggests that between
redshifts 1 and 3, $\alpha$ is smaller than $\alpha_0$ by
$\delta=(-0.72\pm 0.18)\times 10^{-5}$ \cite{murphy,murphy01b,webb01}.
Sensitivity to deviations from zero at this level cannot be reached
with BBN, for both astronomical and nuclear-physics reasons.
Combining this with the large difference in redshifts between BBN
($z\sim10^{9}$) and the observed absorption systems ($1 < z < 3$), it
is doubtful that BBN can shed much light on this claim without a
specific model for variation of $\alpha$.

\acknowledgments

We thank Richard Battye and Marc Kamionkowski for helpful
conversations, and R. B. Wiringa for providing the EM contributions to
the light-nuclide masses.  REL was funded by a PPARC Advanced
Fellowship.  KMN was supported by DoE grant DE-FG03-92-ER40701 and
NASA grant NAG5-11725.


\end{document}